\documentclass[conference]{IEEEtran}
\usepackage{amsmath}
\usepackage{graphicx}
\usepackage[usenames,dvipsnames]{xcolor}
\usepackage[normalem]{ulem}

\begin{document}

\title{Clustering of Pain Dynamics in Sickle Cell Disease \\ from Sparse, Uneven Samples}

\author{\IEEEauthorblockN{Gary K. Nave Jr.\IEEEauthorrefmark{1}\IEEEauthorrefmark{4}, Swati Padhee\IEEEauthorrefmark{2}, Amanuel Alambo\IEEEauthorrefmark{2}, Tanvi Banerjee\IEEEauthorrefmark{2}, Nirmish Shah\IEEEauthorrefmark{3}, and Daniel M. Abrams\IEEEauthorrefmark{1}}
\IEEEauthorblockA{\IEEEauthorrefmark{1}Engineering Science and Applied Mathematics, Northwestern University, Evanston, IL, USA}
\IEEEauthorblockA{\IEEEauthorrefmark{2}Computer Science and Engineering, Wright State University, Dayton, OH, USA}
\IEEEauthorblockA{\IEEEauthorrefmark{3}Department of Medicine, Duke University, Durham, NC, USA}
\IEEEauthorblockA{\IEEEauthorrefmark{4}Email: Gary.Nave@northwestern.edu}}

\maketitle

\begin{abstract}
Irregularly sampled time series data are common in a variety of fields. Many typical methods for drawing insight from data fail in this case. Here we attempt to generalize methods for clustering trajectories to irregularly and sparsely sampled data.  We first construct synthetic data sets, then propose and assess four methods of data alignment to allow for application of spectral clustering.  We also repeat the same process for real data drawn from medical records of patients with sickle cell disease---patients whose subjective experiences of pain were tracked for several months via a mobile app.

We find that different methods for aligning irregularly sampled sparse data sets can lead to different optimal numbers of clusters, even for synthetic data with known properties.  For the case of sickle cell disease, we find that three clusters is a reasonable choice, and these appear to correspond to (1) a low pain group with occasionally acute pain, (2) a group which experiences moderate mean pain that fluctuates often from low to high, and (3) a group that experiences persistent high levels of pain.

Our results may help physicians and patients better understand and manage patients' pain levels over time, and we expect that the methods we develop will apply to a wide range of other data sources in medicine and beyond.
\end{abstract}


%
\IEEEpeerreviewmaketitle

\section{Introduction}
Many biomedical measurements, including pain, are taken at irregular intervals.
Within a hospital environment, for example, patient vital signs are measured based on hospital staff schedule and patient availability \cite{padhee2021pain}.
For self-reported data collected through a mobile app \cite{Jonassaint2015}, the irregularity may be even greater.
Unevenly sampled data present a unique challenge for analysis, because the data do not allow for direct one-to-one comparison of measurements \cite{kreindler2006effects}.
This becomes even more challenging when the sampling is also sparse \cite{liu2016learning}.

If this challenge can be overcome, however, doctors and patients may benefit through a deeper understanding of the different ways in which patients can experience complex diseases \cite{schulam2015clustering} (e.g., by identifying patient subgroups sharing particular dynamics).
In the present study, we consider methods for aligning sparsely and unevenly sampled data into a common coordinate system, applied to both synthetic data and a real data of patients' self-reported pain over time, and seek to identify a classification for the ways that patients with sickle cell disease experience pain.

Other examples of problems with unevenly sampled data comprise topics as varied as the brightness of celestial bodies \cite{naul2018recurrent}, gene expression in cells \cite{moller2005clustering}, bidding in online auctions \cite{peng2008distance}, and paleoclimate reconstruction \cite{rehfeld2011comparison}.
Approaches to handling unevenly sampled data are similarly varied, including recurrent neural networks \cite{naul2018recurrent}, Gaussian processes \cite{liu2016learning}, model-based clustering \cite{james2003clustering}, and frequency-domain analysis \cite{rehfeld2011comparison}.
Here, we focus on ways to align unevenly sampled data into consistent coordinates as a means to generalize standard clustering techniques to unevenly sampled data.

The real-world data considered in the present study measures the long-term pain dynamics of individuals with sickle cell disease.
Sickle cell disease (SCD) is a cardiovascular disease affecting over 5 million people world-wide in which rigid misshapen red blood cells can occlude blood vessels, causing pain and eventual organ damage \cite{rees2010sickle}. 
Pain is the most common symptom associated with SCD, through either acute pain events or through chronic pain.
More than 50\% of patients have at least one hospitalization each year for pain, and 1\% of patients have 6 or more \cite{platt1991pain,wang2010neurobiological}.
Chronic pain is clinically defined as pain on more than half of the days over the previous 6 month period;
about 30\% of adults with SCD are diagnosed with chronic pain \cite{brandow2017sickle}. 
Beyond the classification of chronic pain, there do not currently exist descriptions of the different ways in which patients' pain varies over time.

The SMART app was developed to monitor and interact with patients with sickle cell disease \cite{Jonassaint2015}.
A key feature of the app included self-reporting of pain scores on a scale of 0-10.
Participants were asked to report pain scores twice a day, but actual participation varied significantly.
In all, we consider data from 39 individuals with various data volumes and frequencies; the same data have been used previously to develop a hybrid statistical and mechanistic model for the relationship between pain and medication \cite{clifton2017hybrid}.
Within this data set, the 39 participants reported their pain an average of 67.2 times over an average of 164.6 days.
These pain scores show how patients' experiences with pain changed over time. 
In the present study, we investigate whether it is possible to classify patients' experiences with pain over time based only on their reported pain scores.
This  may be particularly helpful in the future, as promising studies have shown the potential to infer subjective pain from physiological data from wearable devices and electronic health records \cite{padhee2021pain,panaggio2021subjective}.

To discover whether classes of patients with different typical pain dynamics exist, we wish to generalize existing clustering methods to unevenly and sparsely sampled data. 
We propose and compare four different methods for aligning the data into common coordinates: two interpolation-based approaches (nearest-neighbor and linear) and two based on the distribution of reported data.

Unfortunately, existing methods for irregularly sampled are not applicable to our problem.  Approaches based on least-squares spectral analysis (e.g., the Lomb-Scargle (LS) periodogram \cite{scargle1982studies,vanderplas2018understanding}) are oriented toward frequency-domain analysis and can't be applied directly to clustering, especially of sparsely sampled trajectories. 
Model based estimators \cite{harteveld2005estimation} would require us to propose a model for pain dynamics before clustering.

The simple interpolation methods we test have the disadvantage that they require strong assumptions about what happens between measurements of reported pain, but they do permit standard time series analysis techniques.
The distribution methods do not require any assumptions about pain dynamics between measurements (beyond the assumption of a stationary process), but they lose the information contained in the temporal ordering of pain reports.
We compare these four approaches by applying spectral clustering and evaluating the results through a variety of cluster metrics described below.

\begin{figure}[!t]
\centering
\includegraphics{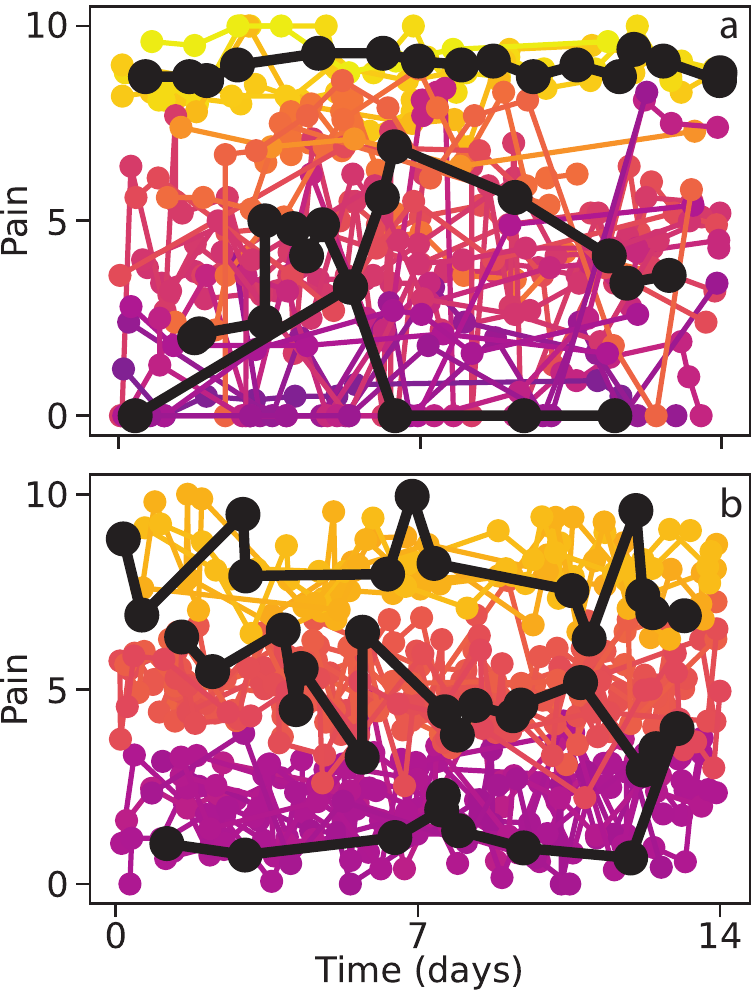}
\caption{Windowed trajectories of pain values for (a) the real data and (b) the synthetic data set.
Three example trajectories are shown in black, while 30 randomly selected examples are shown in the background, colored according to their means.}
\label{fig:raw_trajectories}
\end{figure}

\begin{figure*}[!t]
\centering
\includegraphics{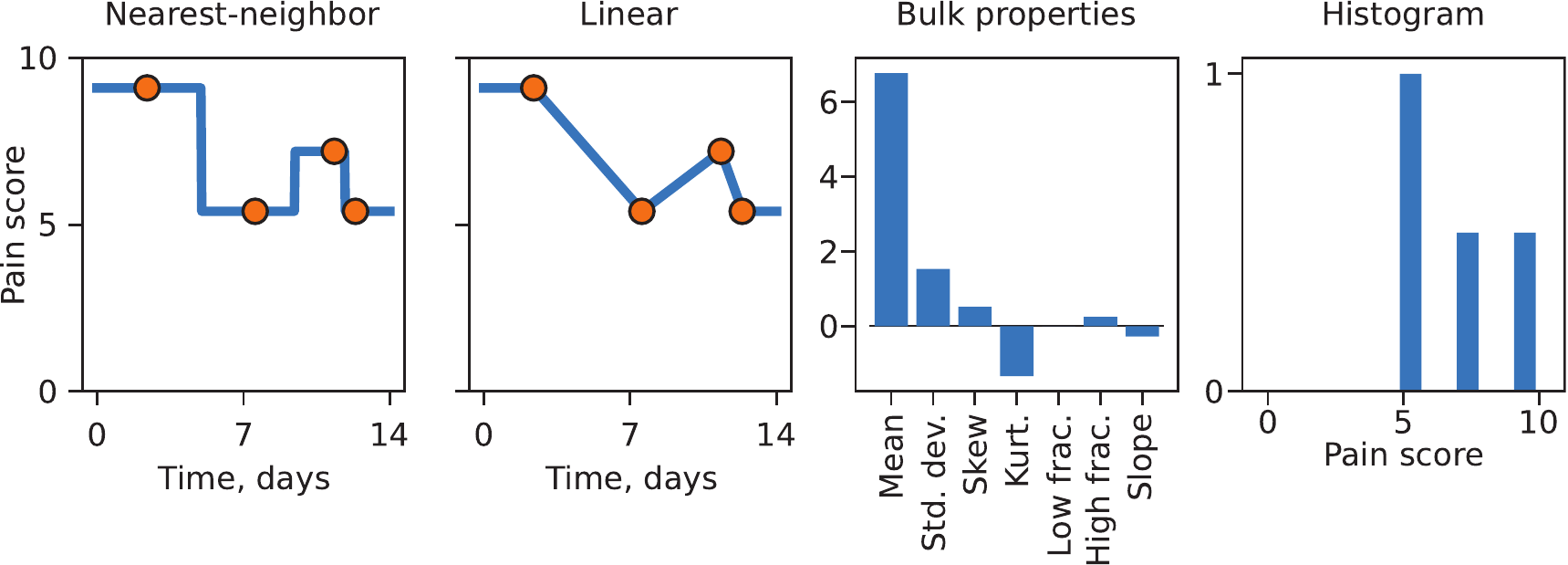}
\caption{A single example trajectory shown transformed into different coordinate systems, as described in Section~\ref{sec:alignment}. This trajectory contains only 4 samples over the 14 day period, shown in orange. The nearest-neighbor and linear interpolations are 672 points, sampled every 30 minutes.}
\label{fig:transforms}
\end{figure*}

\section{Methodology} \label{sec:methods}
To cluster pain trajectories into groups, one needs a measure of distance or similarity between two trajectories, which is particularly challenging when they do not have a consistent number of measurements or times at which measurements are taken. 
This challenge is highlighted in Figure \ref{fig:raw_trajectories}, which shows three example trajectories as they evolve in time.
Most distance metrics rely on data being of a consistent dimension, and so the key step of this work is that of \emph{data alignment}, transforming the data into a common space for similarity measures. 

\subsection{Preprocessing}
We first window the data into two-week periods. 
This windowing was done to increase the number of trajectories and to standardize their lengths. 
The length of 14 days was chosen based on clinical experience to be short enough that pain dynamics are approximately invariant over that time period, but long enough to ensure that acute pain events can be fully captured within a trajectory. 
Windows containing fewer than three pain recordings are removed from the analysis.

We also include synthetic data for comparison with the real data.
This synthetic data set is designed to mimic the irregularly sampled behavior of the real data, but have a predictable clustering structure.
We designed the data set to have three clusters of trajectories centered around pain values of 2, 5, and 8 to provide three distinct groups.
To generate the synthetic data, the number of measurements is first taken from a uniform distribution between 3 and 28.
Then, the timestamp for each of those measurements is drawn from a uniform distribution between 0 and 14 days and the measurements are sorted.
Finally, the pain value at each measurement is drawn from a normal distribution centered around the cluster center value with unit standard deviation.
Values above 10 or below 0 are set to the boundary value. 

\subsection{Data alignment} \label{sec:alignment}
After windowing the data into two week groups, the data are aligned into common coordinates for clustering.
Four different data alignment methods were considered, which are described here. 
These alignment approaches are also highlighted in Figure \ref{fig:transforms}.

In \textbf{Nearest-neighbor interpolation} and \textbf{linear interpolation}, the 14 day window is divided into regularly spaced 30 minute increments, and values between entries are interpolated to either equal the nearest measurement or linearly between measurements, respectively.
Linear interpolation assumes constant values outside of the first and last measurements.
These interpolations are densely sampled to ensure that the data from closely-spaced measurements are not lost.

\textbf{Bulk statistical properties of reported pain levels}: For the first data transformation, we consider properties of the distribution of reported pain levels within each sample. 
Properties considered include the mean, standard deviation, skew, kurtosis, fraction of entries where pain is 0, fraction of entries where pain is high ($\geq 8$), and mean slope of the trajectory.
Each bulk property is then normalized by dividing all entries by the maximum magnitude, so that each property scales from either 0 to 1 or from -1 to 1.
This way, no single property will dominate the distance between trajectories.
Figure \ref{fig:transforms} shows these properties before transformation.

\textbf{Histograms of reported pain levels}: Another approach involves directly comparing the histograms of reported pain values.
The reported pain values are binned into 20 evenly spaced bins and normalized by the total number of reported values within each trajectory.
In calculating the distance between two histograms, we use the cumulative distribution functions.
To understand why, consider two constant trajectories at pain levels of 4 and 5.
Their histograms would each be entirely contained within a single bin, and the distance between the histograms would be large (assuming a norm like $L_1$ or $L_2$), even though the trajectories are very similar.
Their cumulative distributions, on the other hand, would be step functions which increase from 0 to 1 at 4 and 5 respectively.
The distance between these histograms would be small, as desired.

The different transformations of the data provide vastly different representations, and each approach could potentially be justified. 
The transformations also present the data in wildly different dimensionalities, with the bulk properties approach having just 7 dimensions, the histogram approach having 20, and the interpolations representing the data in 672 dimensions.
The histogram approach loses all time information contained in the data, and the bulk properties approach contains time information only in the average slope.
However, the interpolation approaches make strong assumptions about what is happening between pain entries within each sample---assumptions likely (based on clinical experience) to be unjustified.

\begin{figure*}[!t]
    \centering
    \includegraphics{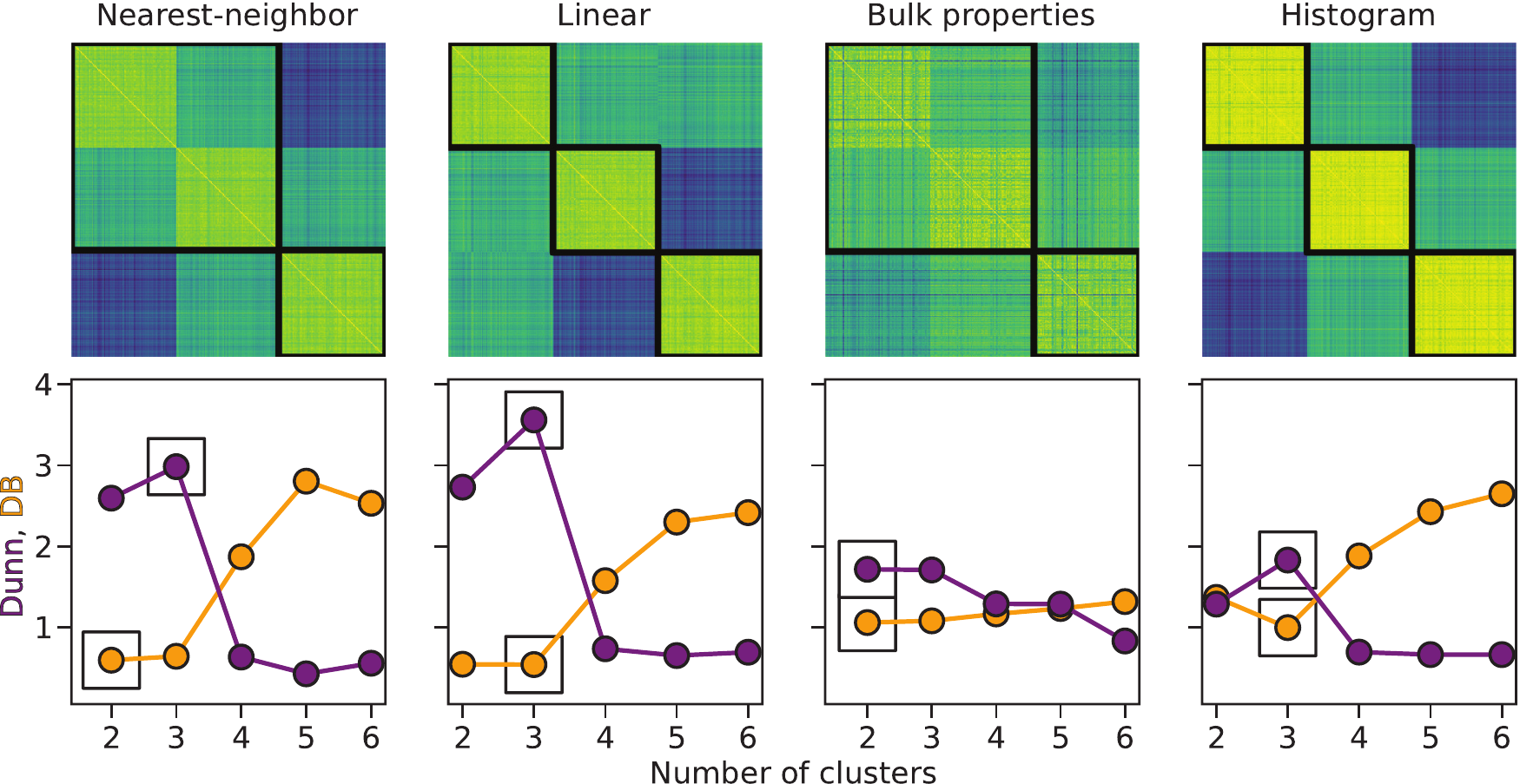}
\caption{Reordered similarity matrices (top) and the Dunn and Davies-Bouldin indices (bottom) for each of the data alignment methods discussed in Section \ref{sec:alignment}, for synthetic irregularly sampled data. The Davies-Bouldin Index is shown in orange, and the Dunn index is shown in purple. Optimal choices for the number of clusters from each index are boxed. The similarity matrices show the optimal clusterings according to the Davies-Bouldin Index, ordered by cluster size.}
    \label{fig:affinities_synth}
    \end{figure*}

\subsection{Clustering}
To cluster the transformed trajectories into groups, we first constructed a matrix of similarities between trajectories, where the similarity, $s$ is defined as,
\begin{equation*}
	s(\mathbf{x}_i, \mathbf{x}_j) = 1 - \frac{d(\mathbf{x}_i, \mathbf{x}_j) - d_{min}}{d_{max}-d_{min}},
\end{equation*}
where $d$ is the distance between points and $d_{min}$ and $d_{max}$ are the minimum and maximum of all distances, respectively.
The Euclidean distance is used for the bulk properties as well as the nearest-neighbor and linear interpolated trajectories.
The Manhattan distance (i.e., the $L_1$ distance, the sum of the absolute values of differences) was applied to the cumulative distribution functions from the histograms in order to calculate the integrated area between curves.
These similarities form a dense, weighted network of trajectories, where nearby trajectories have similarity near 1, and trajectories which are further apart have a similarity approaching 0.

\begin{figure*}[!t]
    \centering
    \includegraphics{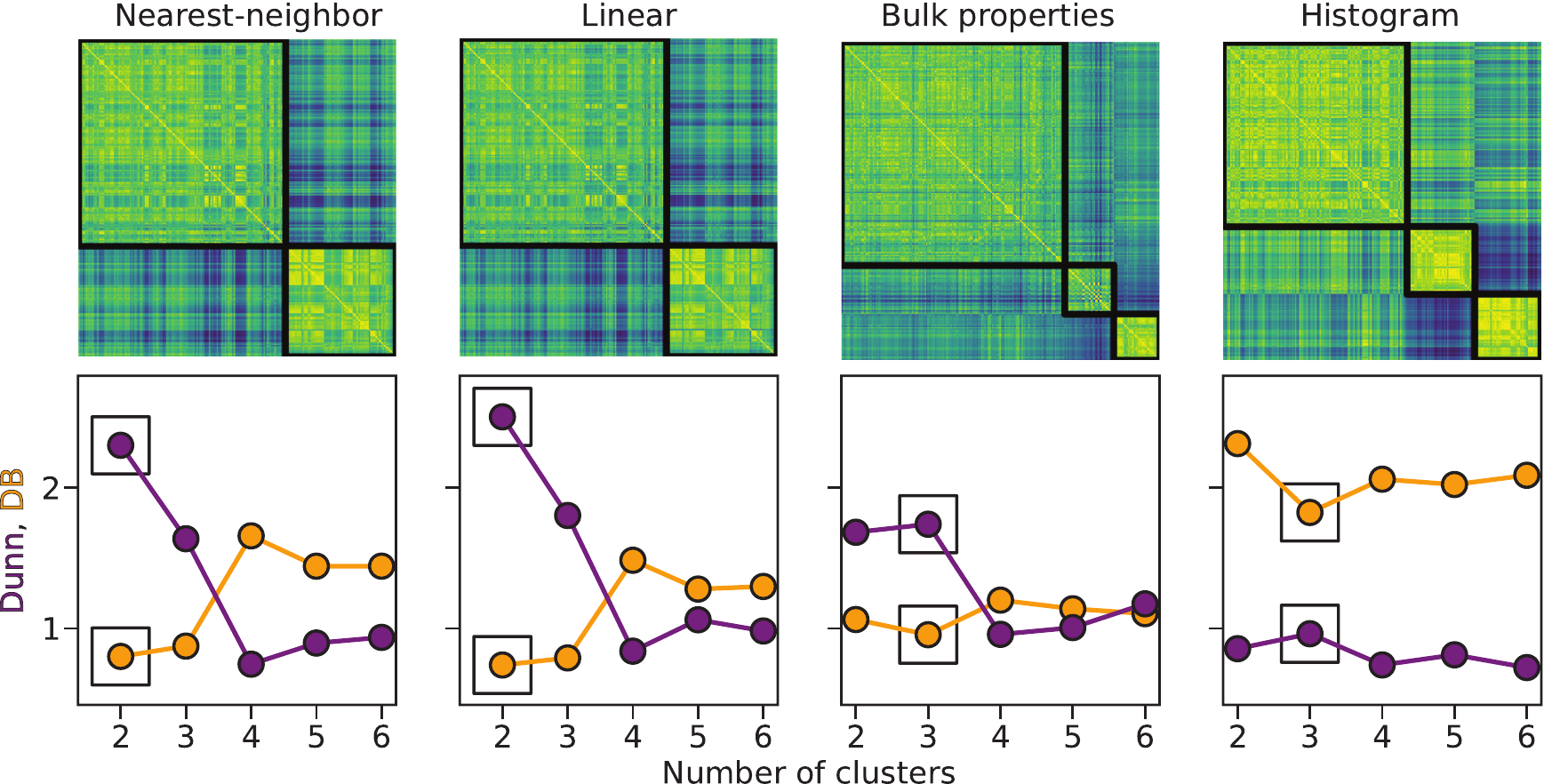}
    \caption{Reordered similarity matrices (top) and the Dunn and Davies-Bouldin indices (bottom) for each of the data alignment methods discussed in Section \ref{sec:alignment}, for patients' self-reported pain data. The Davies-Bouldin Index is shown in orange, and the Dunn index is shown in purple. Optimal choices for the number of clusters from each index are boxed. The similarity matrices show the optimal clusterings according to the Davies-Bouldin Index, ordered by cluster size.}
    \label{fig:affinities}
    \end{figure*}

We apply spectral clustering \cite{von2007tutorial,ng2002spectral} to the normalized similarity matrix between trajectories. 
In spectral clustering, k-means clustering \cite{newman2018networks} is performed on normalized eigenvalues of the graph Laplacian \cite{newman2018networks} of this similarity network, which allows for clusters of different sizes to be more easily identified.
The similarity matrices for the synthetic data and the real data, ordered into clusters, are shown in Figures \ref{fig:affinities_synth} and \ref{fig:affinities}, respectively.

\subsection{Evaluation}
For each data alignment, the clustering process was repeated over an increasing number of specified clusters, and the results of those clusterings were evaluated by the Dunn index \cite{dunn1974well} and the Davies-Bouldin index \cite{davies1979cluster}. 
The Dunn index and Davies-Bouldin index are cluster validity measures used to evaluate the quality of a clustering.

The Dunn index measures the ratio of the minimum distance between cluster centers to the cluster size, given by the maximum mean distance between samples within a single cluster:
\begin{equation}
    DI = \frac{\min_{i,j}D_{ij}}{\max_{i}S_i},
\end{equation}
where $S_i=\tfrac{1}{n_i}\sum_j^{n_i}d(x_j,c_i)$ is the mean distance of points $x$ in cluster $i$ to the cluster center $c_i$, $n_i$ is the number of elements in cluster $i$, and $D_{ij}=d(c_i,c_j)$ is the distance between cluster centers.
A larger Dunn index indicates a better clustering; the distance between clusters is larger than the biggest cluster.

The Davies-Bouldin index is similar, but instead calculates the size of each cluster, $S_i$, and the intercluster distance between each pair of clusters, $D_{ij}$, and then calculates the Davies-Bouldin index $DB$ with,
\begin{equation}
	DB = \frac{1}{K}\sum_{i=1}^K\max_{j\in 1..K}\frac{S_i+S_j}{D_{ij}},
\end{equation}
where $K$ is the number of clusters.
The Davies-Bouldin index gives the worst ratio of cluster size to intercluster distance for each cluster, and then takes the average over all clusters.
A lower Davies-Bouldin index indicates that the cluster sizes are small relative to the intercluster distance for all pairs of clusters, and therefore lower values of $DB$ represent better clusterings.
For the real data, the relevance of clusters we discovered was validated based on the clinical experience of coauthor Shah.

\section{Results} \label{sec:results}
\subsection{Clustering results}
By evaluating the clustering metrics for each data alignment method, we can directly compare them.
Figure \ref{fig:affinities_synth} shows the clustered similarity matrix for the optimal number of clusters according to the Davies-Bouldin index, as well as the Dunn and Davies-Bouldin indices for each number of clusters up to 6 for the synthetic data.
Figure \ref{fig:affinities} shows the same comparison for the real data.

The first notable result from Figure \ref{fig:affinities_synth} is that 3 clusters was not identified to be the optimal number of clusters for all data alignment methods applied to the synthetic data.
Both the bulk properties approach and the nearest-neighbor interpolation identify that the data should be partitioned into two clusters, according to the Davies-Bouldin index.
The Davies-Bouldin index for the nearest-neighbor interpolation of synthetic data is nearly identical for 2 and 3 clusters, so the approach still seems reasonable.
The Dunn index is also highest at 3 clusters for nearest-neighbor interpolation.

However, for the bulk properties approach, neither the similarity matrix nor the Davies-Bouldin index shows the three cluster behavior that we would expect. 
This is likely due to the contrast between the construction of the synthetic data and the statistical properties selected.
Each synthetic sample is given the same standard deviation and is expected to be approximately constant in time, the means of the three clusters are all on the edges of the middle (rather than all the way at the extreme), and the samples are drawn from normal distributions, so we would expect an approximately constant kurtosis and zero skewness. 
Therefore, within the bulk statistical properties selected, the data should mostly vary in the mean, which is just 1 out of 7 dimensions.
The normalization methods used may also have the effect of amplifying noise in this case.  Despite these issues, however, the identification of three clusters is only marginally worse than two according to either the DB or the Dunn index.

The real data (see Figure~\ref{fig:affinities}) also show a disagreement in optimal number of clusters among the data alignment methods.
Both the bulk properties approach and the histogram yield 3 clusters according to both cluster validation measures, although they disagree on the sizes of the clusters.
The interpolation approaches yield 2 clusters.
Both of the interpolation schemes yield only slightly higher Davies-Bouldin indices for 3 clusters than for 2 clusters.
The Dunn index, on the other hand, is clearly higher for 2 clusters than for 3.

\begin{figure*}[!t]
    \centering
    \includegraphics{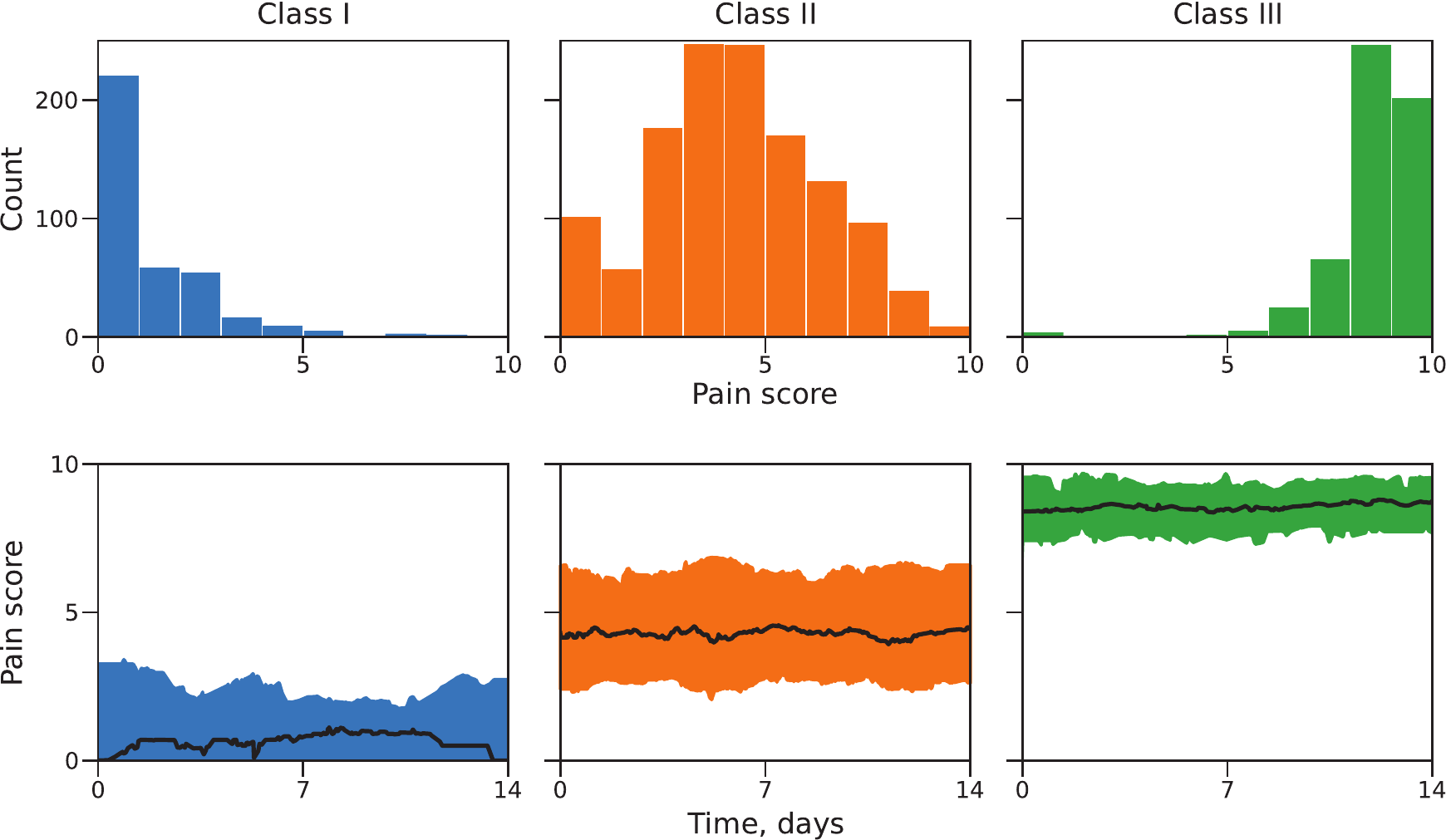}
    \caption{Properties of the clusters identified by the histogram method.
    The top panels show the histograms of all trajectories in each cluster, binned into 10 groups.
    The bottom panels show the time dynamics of each cluster, taken from the linear interpolations of each trajectory.
    The black trajectory shows the median value at each point in time, while the shaded region shows the range from the 16th percentile to the 84th percentile at each point in time.}
    \label{fig:classcomparison}
\end{figure*}

\subsection{Cluster behaviors for real data}\label{sec:clusterbehavior}
To show the lessons learned from clustering, we will select one set of clusters and investigate what the groupings say about the data.
Although there is some ambiguity in the optimal number of clusters, a clustering of 3 groups gives more information than a clustering of 2, allowing for more distinctions between individual trajectories.
Further, even in situations where 2 clusters is optimal, the improvement from 2 over 3 is smaller than the improvement from 3 clusters over 4, so 3 clusters seems to be a reasonable choice.
Here, we will describe in more detail the clusters identified by the histogram method in Figure \ref{fig:affinities}.
Figure \ref{fig:classcomparison} shows the histograms of reported pain data and the time dynamics of the identified classes. 

Class I represents the low pain group, comprising 47 trajectories with a mean of 1.12 and a standard deviation of 1.57. 
Patients in this group tend to report no pain most of the time, with occasional low to medium pain entries.
This class shows 1-2 entries of individuals reporting a very high pain of 8 and 9, as shown in Figure \ref{fig:classcomparison}.

Class II represents the medium pain group.
It is the largest group, with 128 trajectories and a mean and standard deviation of 4.22 and 2.12. 
This group has the highest standard deviation, and individual trajectories often have large variation over time: patients in this group tend to have highly variable pain, usually nonzero.

Class III represents the highest pain group, with a mean and standard deviation of 8.57 and 1.23 from 46 trajectories.
Patients in this group tend to have persistent high levels of pain.
Nearly all measurements lie between 8 and 10.

\subsection{Comparison with chronic pain diagnosis}
Next, we compare the groups themselves with the diagnosis of chronic pain.
For our purposes, this diagnosis is defined by whether or not the patients have been prescribed long-acting opioids. 
Chronic pain patients comprised 22 out of 47 (46.8\%) trajectories in Class I, 107 out of 128 (83.6\%) trajectories in Class II, and 39 out of 46 (84.8\%) trajectories in Class III.

\section{Discussion} \label{sec:discussion}
When data are sparsely and irregularly sampled in time, it can be difficult to make clear decisions about the way to handle clustering.
Here, we present four different methods for transforming the data into a consistent dimensional space, so that trajectories may be directly compared in order to calculate the distance or similarity between trajectories.

The methods we present do not agree on the optimal number of clusters, even for the case of well-separated synthetic data.
The distribution-based approaches lose time information, instead focusing on the distribution of entries, but make no assumptions about what happens between entries.
The interpolation approaches do carry time information and allow for traditional time series methods, but make strong assumptions about what happens between entries which may or may not be accurate.
For instance, it is possible that some patients only chose to enter pain when they took pain medication, which would imply that their pain should be significantly lower between entries.
Unfortunately, unless we apply a specific mechanistic model for pain dynamics, we have no way of knowing what happens between measurements.
Some promising work has been done on developing such mechanistic models \cite{clifton2017hybrid}, but more work remains to be done to establish better models of pain dynamics.

The classes discussed in Section \ref{sec:clusterbehavior} and shown in Figure \ref{fig:classcomparison} give an interesting comparison with the diagnosis of chronic pain. 
Both Class II and Class III have more than 80\% of the trajectories from patients with chronic pain, according to their medication, but the pain dynamics are substantially different.
Within Class III, patients experience persistent extreme pain, with very low standard deviation.
On the other hand, within Class II, patients have a lot of variation in the pain they experience, both within the cluster and within individual trajectories.
These patients generally experience frequent pain, and therefore would fall under the definition of chronic pain.
However, their pain varies a lot from day to day, and is generally not as extreme as the patients within Class III.

A recent study classified patient pain into groups through Latent Dirichlet Analysis of patient survey responses \cite{knisely2021severe}.
In their findings, the authors described three groups: (1) patients with low pain in both the past 7 days and past 6 months, (2) patients with moderate pain over the past 7 days and high pain over the past 6 months, and (3) patients with high pain over both the past 7 days and the past 6 months.
These groups align nicely with the three classes discovered in the present study.
Some patients experience relatively little pain, some patients experience elevated pain levels but not consistently, and some patients experience persistently high pain.

\section{Conclusions} \label{sec:conclusion}
In the present study, we have applied unsupervised learning to the pain dynamics of patients with sickle cell disease, as self-reported by patients using a mobile app.
The nature of data collection led to the challenge of aligning irregularly sampled data. 
We have compared several data alignment methods applied to the pain dynamics data as well as to synthetic data, and shown that the choices made in aligning the data may significantly affect the results of clustering.
For pain associated with sickle cell disease, three clusters appears to be the most reasonable choice, and we have presented a detailed view of the classes discovered.  

More fundamental theoretical work is clearly needed to carefully quantify the robustness and confidence level for proposed clusterings in cases like these, where data is sparse and irregularly sampled. 
We speculate that, given the ever-increasing reserve of medical time-series data, automated grouping of patients into similarity clusters will have great clinical value in the future.

\section*{Acknowledgment}
The authors would like to thank the the National Institutes of Health for support through grant R01AT010413.

\bibliographystyle{IEEEtran}
\bibliography{scdpain}
\end{document}